\documentstyle[aps,twocolumn]{revtex}
\begin{document}
\title{Superconductivity in High $T_c$ Cuprates: The Cause is
No Longer A Mystery}

\author{Philip W. Anderson}
\address {Joseph Henry Laboratories of Physics\\
Princeton University, Princeton, NJ 08544}
\maketitle
\begin{abstract}
 
I discuss various direct calculations of the properties of the 
one-band Hubbard model on a square lattice and conclude that these 
properties sufficiently resemble those of the cuprate superconductors 
that no more complicated interactions are necessary to cause high Tc 
superconductivity..  In particular, I discuss phonon effects and 
conclude that these may be effective in reducing Tc and the gap in 
electron-doped materials.
\end{abstract}

The standard preamble for all kinds of papers on  theory (or 
experiment, for that matter) in the field of high Tc usually contains 
the phrase ``since there is no consensus on the
cause of high Tc superconductivity" or words to that effect, and
often proceeds to justify thereby yet another implausible conjecture
as to some aspect of the phenomenon.  Of course one man's consensus
is another man's wild disagreement, and you can easily find those who
do not agree that there is a consensus on relativity, the quantum
theory, or the Standard Model, by doing a simple web search.  Not, of
course , to mention theories of evolution or of the big bang, which
are unpopular in whole states and most legislatures.  As a solid 
stater I thank my
stars that the band theory of solids and the BCS model are such
obscure targets that they do not garner this kind of disapprobation,
although each had its powerful  though somewhat irrational opponents
in the past--names like Slater and Wigner, even, for the latter.

I would agree that there is or should be no real agreement as to the
cause of much of the peculiar phenomenology of the cuprates.  Each of
us would-be experts has his favorite list of really puzzling
questions about them.  My own favorites are the peculiar
insensitivity of Tc to disorder, and the strange transport
properties in the normal state.  Another  mystery is the role of
interlayer interaction effects in all kinds of ways, although we  can
now be reasonably certain that they do not cause superconductivity.
Yet it is   time , I feel, that it should no longer be legitimate to doubt the
``first cause", namely the minimal underlying model which produces,
among a bewildering welter of other effects, superconductivity at
still unprecedentedly high temperatures.  I can speak with a certain
lack of bias on this matter because the interlayer tunneling  theory
which I advocated for five years, and wrote a book about,  turns out
to be one of those which must be consigned to the dustbin.  (Although
before I was wrong, I was right,  at least partially, as to the
correct ``first cause".\cite{one})

My confidence is based on two things.  First is the rough agreement
between a number of different simulations, extrapolations, and variational
calculations, and the great similarity between these calculations and
the actual experimental data, using a ``bare-bones" model for the
electrons and their interactions, a model for which there is much
independent verification. The second is the experimental data themselves, 
which in many particulars seem to be trying to tell us the nature of the
phenomena.

All of these successful calculations are based on the same simple model for the
underlying physics. This model is the one-band Hubbard model, or, as can be
shown to be essentially equivalent in the appropriate strong-coupling limit,
the $t-J$ model.  It is interesting that there seems to be no need to include
electron-phonon coupling; I will discuss the reason and possible effects of
the phonons later in this article.  The appropriate Hamiltonian, then is
$$
H \ =\ \sum (i|t|j)c* (i)c(j) + U \sum n (i,+) n(i,-) \eqno(1)
$$
$$
H (t-J) \ =\ P \sum (i|t|j)c* (i)c(j)P + \sum J (ij)S(i)* S(j) \eqno(2)
$$
Here $P$ is the projection operator which removes double occupancy of
any site.   (2) is meaningless without the projection.  $+$ and  $-$ of
course are the spin indices.\cite{two}

The most direct calculations with (2) are the quantum Monte-Carlo
simulations of Scalapino and his group,\cite{three} and using a somewhat
different method, those of Sorella and collaborators.\cite{four}  There is
considerable controversy between these two groups as to the details
of the results; however, what strikes the disinterested observer such
as myself is not these differences but the essential similarity of
the two.  The low-temperature phase for low doping is of course
antiferromagnetic ((2) reduces to the Heisenberg model in that case)
but with increasing doping one encounters first an inhomogeneous
(probably striped) phase and then, at approximately the right
temperature and hole density, d-wave superconductivity.  The
parameters of (1) or (2) are not arbitrary--experiments of many types
give us $J$,  and ARPES measurements give us a good estimate of the
band structure at high doping, which is t--so that there is only a
little freedom to manipulate the various components of t.  There are
some confirmatory simulations, particularly on ladder systems, which
again give us a great deal of understanding of why d-wave
superconductivity arises;  there are also strong indications of a
pseudogap phenomenon in the appropriate place.  I am no real expert
on this extremely tricky field, and in fact have been a pessimist
about its ability to arrive at any decisive results, but at this
point I have to confess myself impressed.  Of course, the trouble
with direct calculations is both that they do not really lead to deep
understanding, nor can they say much about such things as excitation
spectra and transport properties, but what they can say is: when you
start out with this physics, you end up with the observed results.  I 
am not at all an expert in the rather involved and subtle reasoning 
that goes into these calculations, so will not present them in 
detail, but there is no question that these are the most extensive 
direct calculations
in existence.

A second approach has been taken by Randeria and Trivedi,\cite{five} by Masao 
Ogata,\cite{six} and to some extent by Sorella's group.  This is to take 
seriously the original ansatz of the RVB theory that the ground state 
wave function could be approximated by Gutzwiller projection of an 
appropriate product wave function of BCS type, and to determine 
variationally the energy gap and the chemical potential of the wave 
function to be projected (which, they point out, need not be the 
physical chemical potential) by calculating the energy and varying 
it.  The Randeria group also improves the wave function somewhat by 
Monte Carlo methods.  Again, as before the results resemble very 
strongly the experimental data insofar as these can be determined 
from the calculation;  in particular, the prediction of d-wave 
superconductivity for an adequately doped sample is very robust. 
Ogata finds that for underdoped concentrations he can find mixtures 
of antiferromagnetism and superconductivity, as is observed in some 
cases.

There is still a third method of direct calculation which has been 
applied, starting with early work by Georges and Yedidia,\cite{seven} and 
carried to an amazing degree of refinement by Bill Putikka\cite{eight} in 
recent years. This is  direct expansion of the  power series in 1/T 
for the partition function, and extrapolation of the results with 
Pade approximants.  This method has played a great role in the past 
in estimating critical behavior near phase transitions, but hasn't 
previously been much used in quantum many-body systems.  It has been 
very useful in disposing of red herrings which have appeared in 
various mean field approximations--showing that the $t-J$ Hamiltonian 
does not lead to phase separation until $J$ is unphysically large, for 
instance; and also identifying the very large $U$, low doping ``Nagaoka" 
ferromagnetic region.  But recently Puttika has carried his series 
out to ten to twelve orders and has been able to calculate n(k) with 
sufficient accuracy to spot the disappearance of the Fermi surface
near the ($\pi$, 0) point which is characteristic of the pseudogap regime. 
The agreement of his results with those of ARPES at the same energy 
resolution is remarkable.

The model is of course not the real substance. Why does it (the 
model) work so well?  In the first place, the question of layer 
interactions:  these are remarkably weak, which we can take as an 
empirical fact on the basis that the Tl one-layer cuprate is 
experimentally seen to exhibit all the phenomena that the others do, 
though proven conclusively by Moler et. al\cite{nine} to have very weak 
interlayer interaction.  The same may be said about the gate-doped 
$CaCuO2$ sample of Kloc, Schon et. al,\cite{ten} where there is only one doped 
layer, but perhaps total acceptance of these results should await 
confirmation on more samples and in other laboratories.  The 
simplification to the simple one-band Hubbard model is often 
justified by the discussion of Zhang and Rice, which is more or less 
correct, but I prefer the use of projective canonical transformations 
a la O K Anderson;\cite{eleven}  and actually the strongest argument was the 
very early exact calculations on small clusters by Michael Schluter 
et. al\cite{twelve} who showed that an effective one-band Hubbard model worked 
very well indeed.

There are two physical aspects which are not in the model and could 
be important. One is long-range Coulomb forces--one assumes effective 
screening as in a normal metal, and the only real justification is 
that it works.  Some experimental data bear on this: we have good 
data on interlayer plasmons which show that the c-axis is a fairly 
insulating direction, actually, (except for supercurrents) but as far 
as we can see the intralayer plasmons are indeed high-frequency and 
may screen reasonably well.

The question which keeps coming up is phonons: why are there almost 
no relevant phonon phenomena?  The answer to this, I would speculate, 
is two-fold.  I believe that it is almost on the level of the 
calculational results above that the nature of the pseudogap is  that 
it is a pairing phenomenon in the spin sector--i e, that the 
pseudogap region can be seen as an RVB with d-wave pairing.  But the 
spin sector is not coupled to phonons in lowest order, because a 
displacement of the local potential does not break the spin 
degeneracy.  In the second term of (2) the electron-phonon 
interaction couples only to J.  It is worth confirming this 
experimentally, but it is known that in heavy-electron materials the 
Kondo spins do not couple to phonons.  Therefore, the pseudogap is 
not affected one way or another by phonons (or ordinary impurities) 
In the point of view that I have called ``RVB redux"\cite{thirteen} the 
superconducting gap is caused by the kinetic energy cost of the 
opening of the pseudogap, which can only be restored by pair hopping 
via the anomalous ``josephson" terms in the kinetic energy;  thus at 
least for the higher doping regime the SC gap  follows the pseudogap.

This is rather a speculative argument, and  I am going immediately to 
say that in fact there are phonon effects on Tc.  For most phonons 
these are reasonably small because, as has been exhaustively proven 
for ordinary superconductors, the electron-electron interaction due 
to phonons is very local and acts primarily between Wannier functions 
on the same site.  In our Hubbard model, it acts simply as a 
modulation of U.  But of course the reason why we have a d-wave is 
that it has zero amplitude at the origin and hence avoids the 
repulsive U, and neither U nor phonons are effective.  But I think 
there is one phonon which can be expected to couple rather strongly 
to our d-electrons, and much more strongly to electrons than to holes 
(note that in our projective theory (2) there is no particle-hole 
symmetry.)  This is the phonon which represents the Jahn-Teller 
displacement which breaks the $d_{x^2 - y^2}$ vs $d_{z^2}$ 
degeneracy (see Fig 1).  An electron site has no ($x^2-y^2$) 
hole and thus tends to relax the Jahn-Teller distortion, while a hole 
site simply adds to the distortion a little.  As is easily seen from 
the figure, the phonon which couples best is at wave 
vector $\pi , \pi$ and hence couples the peaks in the gap function 
 $\pi, 0- > 0, \pi$ This phonon  
has been the subject of considerable discussion recently but rather 
off the point.  What it will do is to cause a repulsive interaction 
for the d-wave gap and hence lower the superconducting Tc for 
electrons relative to holes., an effect which is observed but has 
seemed rather puzzling. It is noteworthy that in a couple of recent 
measurements (see, for instance, [10]) the pseudogap and T* are about 
the same for electron and hole dopings at the same level, but $T_c$ is 
much reduced.

In fact, if the effect is rather localized in k-space we might expect 
the two gaps to behave as in Fig. 2, 
with the $\pi , 0$ peak depressed for  
the SC gap relative to the general point along the Fermi surface. 
This seems to explain the ARPES observations that the peak of the gap 
function is not at the $0, \pi$ points but rather  
there is a local minimum there, as we have sketched in Fig. 2 
(not, of course, to scale)

One implication of this picture is that we would expect an isotope 
effect for the electron-doped materials, but a negative one.  A crude 
calculation, assuming without any justification that the BCS gap 
equation is valid, would suggest

$$
\partial (\ell n \, T_c) / \partial (\ell n \, fopt) \ =\ -\mu/(J-\mu)
\eqno(3)
$$
where $fopt$ is the optical phonon frequency, $\mu$ the phonon coupling 
constant, and $J$ the effective antiferromagnetic exchange .  (3) 
might be valid in the overdoped regime.  For reasons mentioned but 
not emphasized in my book, (which include a great deal of experience) 
I am not very convinced of the accuracy or relevance of isotope 
effect measurements, but perhaps the gate-doping possibility, which 
avoids the necessity of making a new sample for every measurement, is 
a new opportunity.

In conclusion, my point here is that there is a great deal of 
consensus on the  model which underlies high $T_c$ cuprate 
superconductivity, and there ought to be more: I think we have proved 
our point.  But there is much more to be done and specifically, for 
instance, we are so far unable to give a closed-form gap equation.

\vfill\eject

FIGURE CAPTIONS

\begin{enumerate}

\item  {The phonon which is likely to interact strongly with electrons 
around the 0,¼ points.  Its displacements are those of the 
Jahn-Teller distortion caused by an $x^2-y^2$ hole, so that it might be 
expected to be particularly soft for electron doping.
}
\item  {The reduction in the superconducting gap which might result 
from the repulsive interaction caused by the optical phonon of Fig. 1.
}
\end{enumerate}
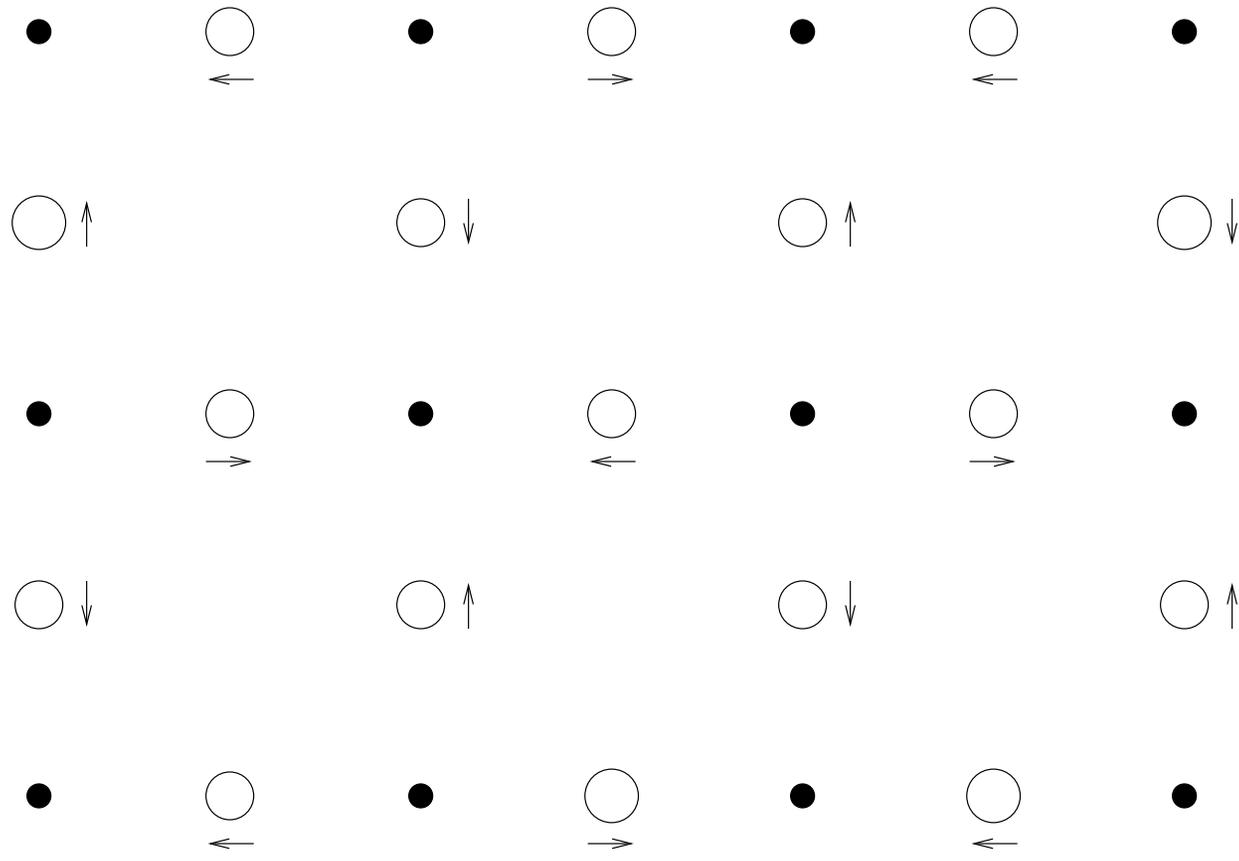
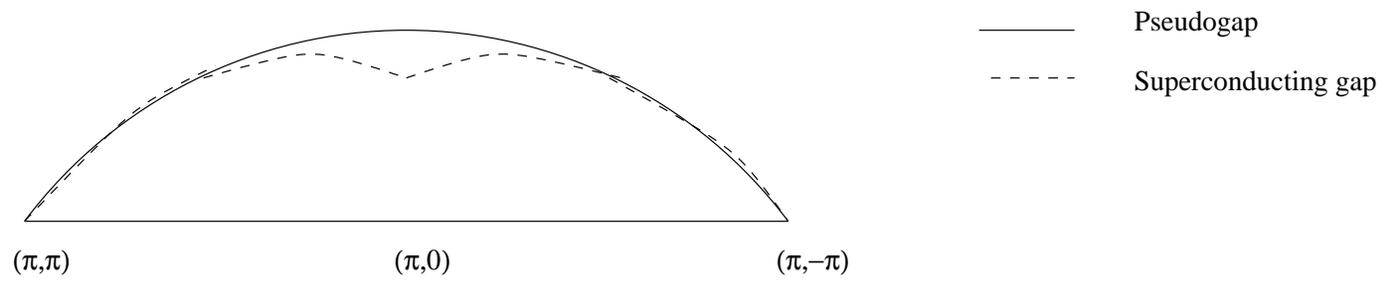
\end{document}